\documentclass[useAMS,usenatbib,usegraphicx]{mn2e}

\title[Two-component outer ring in the Galaxy] {Two-component outer ring and the Galactic spiral structure}
\author[A.M. Mel'nik and  P. Rautiainen]{A.M. Mel'nik$^{1}$\thanks{E-mail: anna@sai.msu.ru}
  and P. Rautiainen$^{2}$\\ $^{1}$Sternberg Astronomical Institute,
  13, Universitetskii pr., Moscow, 119992, Russia\\ $^{2}$Astronomy
  Division, Department of Physics, University of Oulu, P.O.
  Box 3000,\\ FI-90014 Oulun yliopisto, Finland}
\begin{document}

\date{Accepted 2011 December 00. Received 2011 December 00; in original form 2011 December 00}


\maketitle

\label{firstpage}

\begin{abstract}

Model of the Galaxy with the ring $R_1R_2'$ can explain some
large-scale morphological features of the Galactic spiral
structure.  The Carina-Sagittarius arm can  consist of two
ascending segments of the  outer rings $R_1$ and $R_2$ which
almost touch each other near the Carina region.  The Perseus and
Crux arms  can be partially identified with  the descending
segments of the ring $R_2$. Model of the two-component  outer
ring can also explain the existence of some maxima in diagrams
(l, $V_{LSR}$) which are supposed to correspond to the directions
tangential to the spiral arms.    On the basis of numerical
simulations we propose two sketches of the  ring structure of the
Galaxy which include  the bar, two outer rings, the inner ring,
and the nuclear gas condensation, that may be a nuclear ring.
Both sketches can explain the position of the Carina-Sagittarius
arm with respect to the Sun.
\end{abstract}


\section{Introduction}

The best tracers of the Galactic spiral structure are HII regions
-- gas clouds ionized by young hot stars. Their radio emission
penetrates the interstellar dust and they can be observed even in
the distant parts of the Galactic disk. Heliocentric distances
$r$ for the faraway HII regions ($r>6$ êïê) are usually
determined from the kinematical models under the assumption that
velocity deviations from the rotation curve   are zero. The
kinematic method yields  an unambiguous distance for objects
located outside the solar circle ($R>R_0$),  but gives two
possible distances corresponding to the same line-of-sight
velocity inside the solar circle ($R<R_0$, where $R$ -- is the
Galactocentric distance). The choice between ``near'' and ``far''
distances requires additional information, usually it is the data
on the absorption/emission lines of HI/H$_2$CO or self-absorption
in the HI line. The method is based on the analysis of velocities
of the foreground clouds \citep{anderson2009}.

\citet{georgelin1976} using the distribution of 100 HII regions
with the excitation parameter more than $U>70$ pc cm$^{-2}$ have
proposed a 4-armed spiral pattern with the mean pitch angle of
spiral arms of $i\approx 12^\circ$.  Their model can also explain
the existence of so-called tangential directions -- lines of
sight corresponding to maxima in the thermal radio continuum, HI
and CO emission -- which are associated with the tangents to the
spiral arms.  These directions were first determined from the
analysis of the longitude-velocity diagrams in HI
\citep{kerr1970,burton1970,simonson1970} which  exhibited  the
distribution of gas temperature in coordinates (l, $V_{LSR}$) (l
-- the Galactic longitude, $V_{LSR}$ -- the heliocentric
line-of-sight velocity $V_r$ corrected for the solar motion to
the apex) averaged over some range of  Galactic latitudes $b$.
The original model by \citet{georgelin1976} has been developed on
the basis of new data \citep{lockman1979, downes1980,
caswell1987, watson2003, russeil2003,
 paladini2004, russeil2007, hou2009, efremov2011}.

\citet{russeil2003} has grouped HII regions and molecular clouds
into complexes of star formation which enables her to decrease
the random errors in determination of mean velocities and
kinematical distances.  Locations of spiral arms supposed by
\citet{russeil2003} practically coincide with those obtained by
\citet{georgelin1976}, though the spiral structure generally
becomes more symmetrical. \citet{russeil2003}  supposes that her
sample of complexes including HII regions with high excitation
parameter ($U>60$ pc cm$^{-2}$) is complete all over the Galactic
disk. For determination of kinematical distances she has used
nearly flat rotation curve derived from objects with known
photometric distances.

There are also other indicators of the Galactic spiral structure.
One of them  is the giant clouds of molecular hydrogen (GMC) with
the size of $\sim 40$ pc and the mass of 10$^4$ -- 10$^6$
M$_\odot$. \citet{cohen1986} showed that GMCs outlined well the
Carina arm. \citet{dame1986} solved the ambiguity in the choice
from  the two kinematical distances in the first quadrant and
selected objects of the Sagittarius arm. \citet{grabelsky1988}
compiled a catalogue of GMCs in the region $270<l<300^\circ$ and
identified the objects of the Carina arm.  Also, the neutral
hydrogen concentrates on the spiral arms \citep{oort1958,
kerr1962} and is distributed quite non-uniformly outside the
solar circle \citep{henderson1982,kalberla2005, levine2006}.

We will show that two-component outer ring of  class $R_1R_2$ can
also explain many large-scale morphological features of the
Galactic spiral structure. The paper has the following structure:
Section 2 is devoted to tangential directions, the dynamical and
kinematical aspects of the problem are discussed in Section 3, a
brief description of dynamical models including the outer rings
is given in Section 4, Section 5 presents  results of a
comparison of our models with observations.

\section{Tangential directions and  the names of the spiral
arms}

\begin{figure*}
\centering \resizebox{14 cm}{!}{\includegraphics{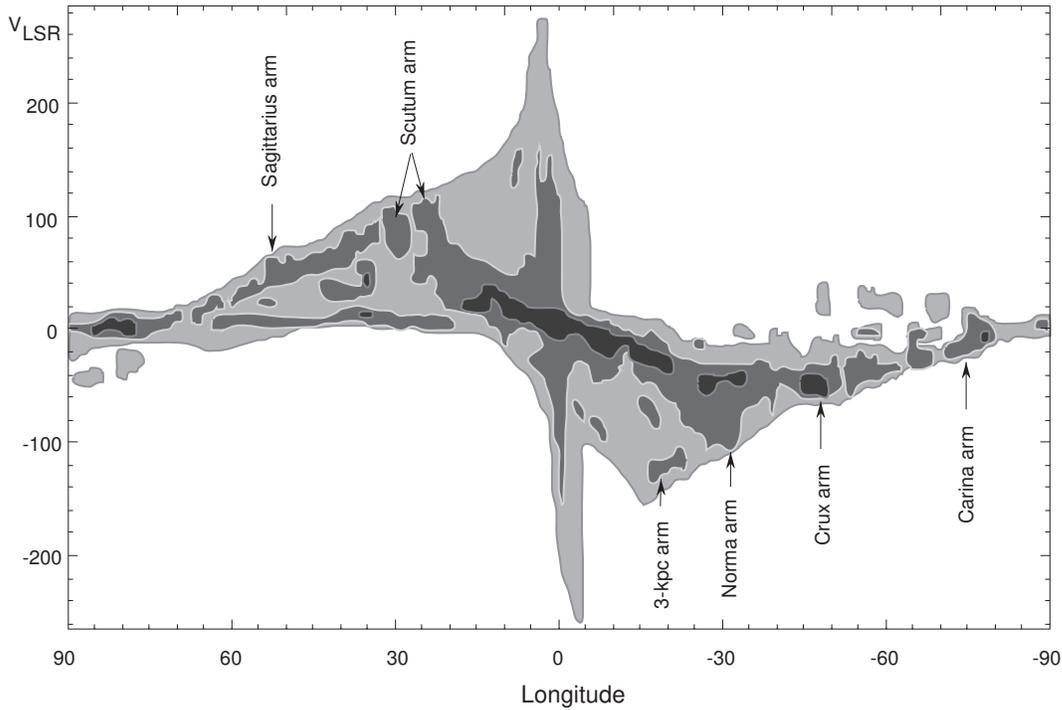}}
\caption{Sketch of the diagram (l,$V_{LSR}$) of  the $^{12}$CO
distribution \citep{dame2001}.  The emission is averaged in the
range $b=\pm2^\circ$. It also indicates the positions of maxima
corresponding to the directions tangential to the spiral arms.}
\label{dame}
\end{figure*}

\citet{englmaier1999} and \citet{vallee2008} compiled information
about directions tangential to the spiral arms. Generally, the
tangential directions are connected with the existence of some
intensity maxima in diagrams (l, $V_{LSR}$). Fig.~\ref{dame}
shows the distribution of $^{12}$CO composed by \citet{dame2001}.
The velocities of more than $\pm 150$ km s$^{-1}$ in the central
region ($|l| \leq 10^\circ$) can be explained by the presence of
elliptical orbits in the central region. But, in general, gas at
the positive longitudes, $10<l<90^\circ$, has the positive
velocities $V_{LSR}$ while at negative longitudes,
$-90<l<-10^\circ$,  the negative ones. The extreme velocities at
each direction are often called terminal velocities. Besides, the
diagrams demonstrate the ridge-like intensity maxima that are
often associated with the spiral arms. The directions where the
``ridges'' reach the curves of terminal velocities are thought as
the tangential directions (Table~1).

\begin{table}
  \caption{Directions tangential  to the spiral arms}
  \begin{tabular}{llll}
   \hline
N& longitude& Name & Other name\\
 \hline
1 & $l\sim284^\circ$  & Carina arm &\\
 &&&\\
2 & $l\sim310^\circ$  & Crux arm & Centaurus arm\\
  &&&\\
3 & $l\sim327^\circ$  & Norma arm & Norma-3-kpc arm\\
 &&&\\
4 & $l\sim339^\circ$ & 3-kpc arm & start of Perseus arm\\
 &&&\\
5 & $l\sim25$, $31^\circ$ & Scutum arm &\\
 &&&\\
6 & $l\sim51^\circ$ & Sagittarius arm & \\
 \hline
\end{tabular}
\end{table}

Note also the presence of the bright emission at $l=80^\circ$
corresponding to the Cygnus region ($l=73\textrm{--}78^\circ$,
$r=1.5$ kpc) which is usually directly associated with the Local
arm or spur (its another name is Orion-Cygnus arm) and is
excluded from consideration.

The connection of  bright spots in the diagrams (l, $V_{LSR}$)
with a certain distance should be taken with great caution: in
reality they can consist of a chain of clouds extended to several
kpc along the line of sight \citep{adler1992}. The problem is
that the different models of the gas motion in the Galaxy can
produce the very similar diagrams (l, $V_{LSR}$).

\begin{figure}
\resizebox{\hsize}{!}{\includegraphics{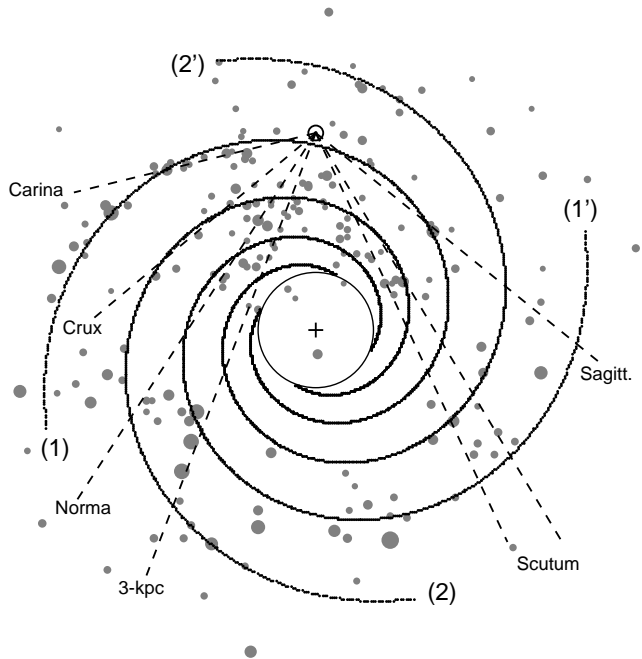}}
\caption{Regular spiral pattern with parameters of logarithmic
spirals: $i=12.8^\circ$, $r_0=2.1$ kpc, $\theta_0=-20^\circ$, and
$m=4$ \citep{vallee2008}. 1: Sagittarius-Carina arm,  2:
Scutum-Crux arm,  $1'$: Norma-Cygnus arm, and $2'$: Perseus arm.
It also shows the tangential directions to the spiral arms. Giant
star-forming complexes ($U>60$ pc cm$^{-2}$) from the catalogue
by \citet{russeil2007} are depicted by grey circles  whose size
is proportional to the excitation parameter.} \label{4-armed}
\end{figure}

Fig.~\ref{4-armed} illustrates the idea of tangential directions.
It shows a regular spiral pattern with parameters:
$i=12.8^\circ$, $r_0=2.1$ kpc, $\theta_0=-20$, and $m=4$ taken
from the paper by \citet{vallee2008}, as well as the tangential
directions. It also exhibits  and distribution of giant
star-forming complexes from the catalogue by \citet{russeil2003}.
We can  clearly see that every ray is tangent (or passes very
close to the tangent) to the spiral arm. On the other hand, only
the Carina arm is outlined well by star-forming complexes.

Note that the  naming of the arms in literature is somewhat
confusing: the Norma arm is sometimes called Norma-3-kpc arm, but
the 3-kpc arm, in its turn, is also termed ``start of Perseus
arm'' (see also Table 1). Another example is the Cygnus arm,
which can easily be confused with the Cygnus region situated near
the Sun ($r=1.5$ kpc). This  outer arm is also sometimes called
``Perseus + I arm'' or ``Norma-Cygnus arm''
\citep{vallee2005,vallee2008}.

There are no tangential directions to the  outer Cygnus arm
($70<l<220^\circ$, $r=5\textrm{--}9$ kpc, $R=11\textrm{--}15$
kpc), because it lies outside the solar circle. Interestingly,
the Cygnus arm is absent on the schema supposed by
\citet{georgelin1976}. Its appearance is caused by two reasons:
the principle of symmetry and  discovery of new HII regions.
\citet{efremov1998,efremov2011} identifies  the HI superclouds
outlining the Carina-Sagittarius arm and shows that the arm
symmetrical to it doesn't coincide with the Perseus arm but lies
beyond it. Additionally, Russeil \citep{russeil2003, russeil2007}
discoveres many star-forming complexes in the region
$70<l<220^\circ$ at the distance range $r=4\textrm{--}10$ kpc
which cannot belong to the Perseus arm.


\section{Dynamical and kinematical  aspects of the problem}

Model suggested by \citet{georgelin1976} and developed in
subsequent papers leaves  open many questions.  At the moment no
N-body simulation with realistic rotation curve and the size of
the bar can  reproduce the classical 4-armed pattern. The main
problem  concerns the  dynamical mechanism which could support
the spiral pattern occupying a large part of the galactic disk
\citep[see surveys by][]{toomre1977, athanassoula1984,
binney2008}.

\begin{figure}
\centering \resizebox{6 cm}{!}{\includegraphics{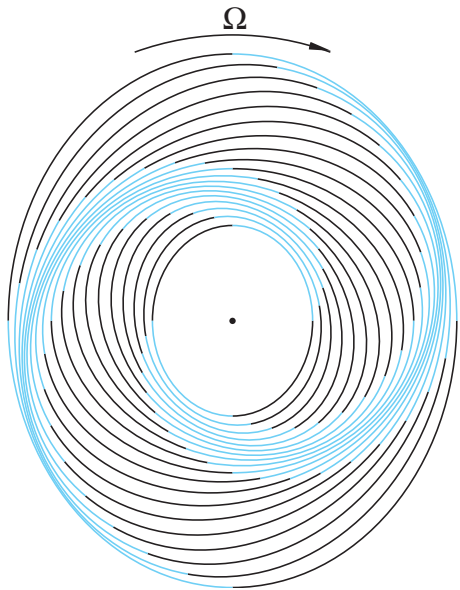}}
\caption{Segments of  stellar orbits with the negative and
positive radial velocity $V_R$ in the trailing density-wave
spiral arms located inside the CR. The galaxy rotates clockwise.
Motions are considered in the reference frame co-rotating with
the spiral pattern, in which stars also rotate clockwise. The
ascending segments of ellipses ($V_R>0$) are shown in black
whereas the descending ones ($V_R<0$) are shown in blue. Inside
the CR the crowding of the orbits occurs on the descending
segments of ellipses.} \label{ellipses}
\end{figure}

The concept of the density-wave theory \citep{lin1964,
bertin1996}  where the spiral arms are forming at places of
crowding of the orbits deserves special attention. A lot of
researches think that at least two major spiral arms in the
Galaxy  are the density-wave spiral arms. But density-waves
create specific distribution of velocities  in the young disk
population that  is forming due to adjustment of epicyclic
motions of stars in accordance with orbital rotation
\citep{lin1969}. \citet{kalnajs1973} suggests to consider stellar
orbits in the reference frame co-rotating with the speed of the
spiral pattern $\Omega_{p}$, in which the orbits are looking as
pure ellipses or ellipses with the ``dimples''. If we know the
direction of rotation of disk stars in the adopted frame then we
can divide their orbital ellipses into the ascending and
descending segments, where stars go away ($V_R>0$) and toward
($V_R<0$) the Galactic center, respectively. Fig.~\ref{ellipses}
illustrates the idea that orbit crowding occurs at the descending
or ascending segments of ellipses, and the choice between them
depends on the sense of orbital rotation. This  must be  viewed
in the reference frame rotating with the speed $\Omega_{p}$ in
which the sense  of rotation is determined by the position of the
corotation radius (CR) with respect to the region considered.
Thus, the knowledge of the direction of radial component $V_R$ of
velocities  in the spiral arms allows us to restrict the region
where the CR can be located and thereby roughly estimate the
value of the angular speed of the spiral pattern $\Omega_{p}$.

The study of the kinematics of  young stars in the regions of
intense star formation  yields unexpected distribution of the
velocities. The radial component $V_R$ of the  velocity  in the
Carina, Cygnus, and Perseus regions is directed toward the
Galactic center ($V_R<0$) while it is directed away from it
($V_R>0$) in the Sagittarius region and in the Local System
\citep{melnik1999,melnik2001,melnik2003, sitnik2003,
melnikdambis2009}.  This means that the Perseus and Sagittarius
regions cannot be parts of the same density-wave spiral pattern
rotating with  one pattern speed. Note that two-armed model of
the Galactic spiral structure with the angular speed of
$\Omega_p=13.5$ km s$^{-1}$ kpc$^{-1}$ \citep{lin1969} can
reproduce well the kinematics in the Perseus region \citep[][and
other papers]{roberts1972,burton1974, humphreys1976}. But the
kinematics of the Sagittarius region indicates that it  must be
located outside the CR and rotates with the speed of more than
$\Omega_p>38$ km s$^{-1}$ kpc$^{-1}$ \citep{melnik2006}.

There are a lot of evidences that our Galaxy includes the bar.
The estimation of the length of the  bar has increased from
initial $R_{bar}= 2-3$ kpc \citep{blitz1991,binney1991,
blitz1993} to the current values  $R_{bar}=3-5$ kpc
\citep{benjamin2005, babusiaux2005,habing2006,cabrera-lavers2007,
pohl2008,gerhard2011}. Dynamical models of the gaseous medium
moving in the Galactic potential perturbed by the bar reproduce
the so-called ``parallelograms'' on the diagrams (l, $V_{LSR}$)
in the central region \citep{weiner1999,fux1999, englmaier1999,
englmaier2006}. The general consensus is that  the major axis of
the bar is oriented in the direction
$\theta_b=15\textrm{--}45^\circ$ in such a way that the end of
the bar closest to the Sun lies in the first quadrant.

The concept  that the Galaxy can include several modes rotating
with different angular speeds was actively developed in the
beginning of the 2000s. The rapidly rotating bar ($\Omega_b=
40\textrm{--}60 $ km s$^{-1}$ kpc$^{-1}$)  and the slower mode
($\Omega_{sp}= 20\textrm{--}40$ km s$^{-1}$ kpc$^{-1}$) could
explain the gas kinematics in the central region and at larger
distances, respectively \citep{bissantz2002,bissantz2003}.
However, application of a two-mode model to the Galaxy   appears
to be much harder than expected. On the one hand, there are many
dynamical models, where the disk forms a pattern rotating slower
than the bar \citep{sellwood1988, masset1997, rautiainen1999,
rautiainen2000}. On the other hand, after introducing  physical
units the strongest slow mode turns to have  the pattern speed of
$\Omega_{sp}\approx 30$ km s$^{-1}$ kpc$^{-1}$, which is too high
to explain the  kinematics of young stars in the Perseus region.

In parallel with the concept of modes a different approach has
been developed. Here the spiral arms are regarded as a subsequent
generation of short-lived spiral perturbations connected with
each other through the resonances: the CR of each next wave is
located at one of the resonances of the previous one
\citep{sellwood1989,sellwood1991,sellwood2000,sellwood2011}.
Nevertheless, it is questionable whether this approach can
explain the existence of long spiral arms similar to the Carina
one in the Galaxy (Fig.~\ref{4-armed}).

\section{Models of the Galaxy including the outer ring}

The essential characteristic of the galaxies with the outer rings
and pseudorings -- incomplete rings made up of spiral arms -- is
the presence of the bar \citep{buta1995,buta1996}. Since the
outer rings have an elliptic form, the broken outer rings
(pseudorings) resemble two tightly wound spiral arms. Two main
classes of the outer rings and pseudorings have been identified:
the $R_1$ rings ($R'_1$ pseudorings) elongated perpendicular to
the bar and the $R_2$ rings ($R'_2$ pseudorings) elongated
parallel to the bar. In addition, there is a combined
morphological type $R_1R_2'$ which shows elements of both
classes. The $R_2$ rings have elliptical shape, but the $R_1$
rings are often ``dimpled'' near the bar ends \citep{buta1995,
buta1991}.

The test particle simulations \citep{schwarz1981, byrd1994,
rautiainen1999} and N-body simulations \citep{rautiainen2000}
show that the outer rings   are typically located in the region
of the Outer Lindblad Resonance (OLR). \citet{schwarz1981}
connected two main types of the outer rings with two main
families of periodic orbits existing near the OLR of the bar
\citep{contopoulos1980,contopoulos1989}. The stability of orbits
enables gas clouds to follow them for a long time period. The
$R_1$-rings are supported by $x_1(2)$-orbits \citep[using the
nomenclature of][]{contopoulos1989} lying inside the OLR and
elongated perpendicular to the bar, while the $R_2$-rings are
supported by $x_1(1)$-orbits situated a bit outside the OLR and
elongated along the bar.

The bar semi-major axis in the Galaxy    is supposed to lie in
the range $a=3\textrm{--}5$ kpc.  For the flat rotation curve and
a fast rotating bar this means that the bar angular speed
$\Omega_b$ is limited by  the interval $\Omega_b=40\textrm{--}70$
km s$^{-1}$ kpc$^{-1}$ and the OLR of the bar is located in the
solar vicinity: $|R_{OLR}-R_0|<1.5$ kpc.   The studies of the
kinematics of old disk stars in a small solar vicinity, $r<250$
pc, revealed a bimodality  in the distribution of ($u$, $v$)
velocities  which was also interpreted as a result of the solar
location  near the OLR of the bar \citep[][and other
papers]{kalnajs1991, dehnen2000, fux2001, chakrabarty2007,
minchev2010}. Thus, the presence of an outer ring in the Galaxy
is a plausible possibility to be considered.

In addition to the outer rings, the Galaxy can include an inner
ring or pseudoring surrounding the bar  which manifests itself in
the so-called 3-kpc arm(s) \citep{fux1999,dame2008,
churchwell2009}. Also,  a hypothesis about the presence of a
nuclear ring with a major axis of $\sim1$ kpc is considered
\citep{rodriguez-fernandez2008}.

With using the simulation code developed by H. Salo
\citep{salo1991,salo2000} we have constructed two different types
of models (models with analytical bars and N-body simulations)
which reproduce the kinematics of OB-associations in the Perseus
and Sagittarius regions. The kinematics of young stars in the
Perseus region indicates the existence of the $R_2$ ring while
the velocities in the Sagittarius region suggest the presence of
the $R_1$ ring in the Galaxy. Our models have nearly flat
rotation curves. The major and minor axes of the bar have the
values of $a=4.0$ and $b=1.2$ kpc. The value of the solar
position angle $\theta_b$ providing the best agreement between
the model and observed velocities is $\theta_b=45\pm5^\circ$. The
bar angular speed lies in the range $\Omega_b=42\textrm{--}55$ km
s$^{-1}$ kpc$^{-1}$ \citep[][hereafter Papers I and II,
respectively]{melnikrautiainen2009, rautiainen2010}.

In the present paper we  use the distribution of OB-particles in
model No. 3 obtained in  series of models with analytical bars
for the time moment $T=15$ ($\sim 1$ Gyr). Model 3 was chosen due
to presence of the inner ring which  still persists by $T= 1$
Gyr. As for the outer rings, all models considered produce the
similar distribution of OB-particles on the galactic periphery
(Paper I). We  also use the distribution of gas and stellar
particles in N-body model averaged  for the time interval
$T=5\textrm{--}6$ Gyr. The averaging over large time interval
reduces the influence of slow modes and occasional perturbations
(Paper II).

\section{Results}

\subsection{Ring $R_1R_2'$ and the distribution of giant
star-forming complexes}

In this section we  will use data from the catalogue  by
\citet{russeil2007}, particularly, the sample of giant
star-forming complexes with the excitation parameter of more than
$U>60$ pc cm$^{-2}$ that includes 194 regions in the range of
Galactocentric distances $0<R<12$ kpc, 76\% of them have only
kinematical distances.

The distance scale in our models (Papers I and II) is adjusted to
the so-called short distance scale of classical Cepheids
\citep{berdnikov2000}. Distance scale for star-forming complexes
from the catalogue  by \citet{russeil2007} $r_0$ is close to that
for OB-associations \citep{humphreys1984,humphreys1989}, so to
match it with the short distance scale, we used the same scaling
factor of $f=0.8$  ($r=fr_0$), which was used for reducing the
distance scale for OB-associations \citep{sitnik1996, dambis2001,
melnikdambis2009}.

\begin{figure*}
\resizebox{\hsize}{!}{\includegraphics{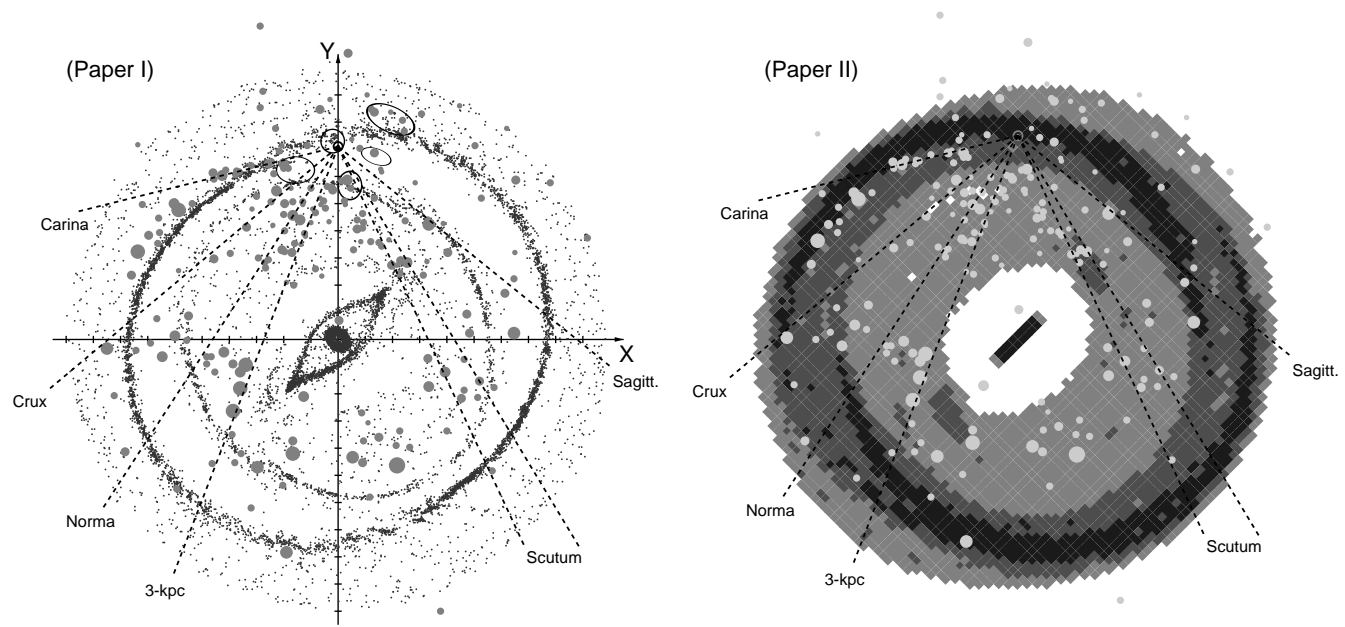}}
\caption{Comparison of the distribution of model particles with
the distribution of giant star-forming complexes
\citep{russeil2007}.  Left panel: distribution of OB-particles
(black points) in model with analytical bar (Paper I). It also
indicates the boundaries of the regions of intense star formation
studied in Papers I and II. A division on the $X$ and $Y$ axis
corresponds to 1 kpc. Right panel: gas-density distribution in
N-body model averaged in small squares (Paper II). The
light-gray, dark-gray and black colors represent squares
containing the increasing number of particles. Giant complexes
($U>60$ pc cm$^{-2}$) are depicted as circles whose size is
proportional to the excitation parameter $U$. The tangential
directions to the spiral arms are also shown (Table 1). The
adopted value for the solar position angle with respect to the
bar is $\theta_b=45^\circ$.  The Galaxy rotates clockwise.}
\label{compare-I}\end{figure*}

Fig.~\ref{compare-I} exhibits the distribution of giant
star-forming complexes  and that of OB-particles from the series
of models with analytical bars (Paper I). It also demonstrates
the position of the regions of intense star-formation studied in
Papers I and II. The Sagittarius region ($x=0.5$, $y=6.0$ kpc)
lies on the segment of the ring $R_1$, whereas the Carina region
($x=-1.5$, $y=6.5$ kpc) occupies the intermediate position
between two outer rings in the place where they come closest to
each other. The Perseus region ($x=2.0$, $y=8.0$ kpc) and the
Local System ($x=0.0$, $y=7.4$ kpc) belong to the ring $R_2$,
while the Cygnus region ($x=1.5$, $y=6.9$ kpc) appears to lie in
the inter-ring space. The Galactocentric distance of the Sun is
adopted to be $R_0=7.1$ \citep{rastorguev1994, dambis1995,
glushkova1998}.

 Outer rings can be divided onto the ascending ($V_R>0$) and
descending ($V_R<0$) segments. On the ascending segments
(segments C-D-E and 5-6-7 in fig.~6 of Paper I), the
Galactocentric distance $R$ decreases with the increase of the
azimuthal angle $\theta$. This becomes clear if we remember that
the closed orbits emerge only in the reference frame co-rotating
with the bar. The outer rings lie near the OLR of the bar where
disk objects rotate slower than the bar, therefore, in the
reference frame co-rotating with it they will move in the
direction opposite that of the Galactic rotation, i.e.
counterclockwise. On the descending segments of the outer rings
(segments 3-4-5 and E-F-G in fig.~6 of Paper I) the
Galactocentric distance $R$ increases with the increase of
$\theta$. Note also that ascending segments of the outer rings
can be regarded as fragments of the trailing spiral arms while
the descending ones -- as those of the leading spiral arms.

The Carina arm  is often regarded as the major spiral arm in the
Galaxy. It begins near the Carina region and unwinds
counterclockwise along the Galactocentric angle at $|\Delta
\theta| \approx 90^\circ$. It is evident that the star-forming
complexes related to the Carina arm  fall nicely on the ascending
segment of the ring $R_2$: the deviation doesn't exceed 15\% of
the heliocentric distance $r$. Note also that the objects related
to the Sagittarius arm are situated near the ascending segment of
the ring $R_1$. Although most of the researchers consider the
Carina-Sagittarius arm as a single spiral arm, it could consist
of two ascending segments of the outer rings $R_1$ and $R_2$
which almost touch each other near the Carina region ($x=-1.5$,
$y=6.5$ kpc). It is difficult to say anything about another pair
of the ascending segments of the outer rings, but it is possible
that they can be identified with the Norma-Cygnus arm symmetrical
to the Carina-Sagittarius one. If the ascending segments of the
outer rings were much brighter than the descending ones, then the
Galactic spiral structure would be considered as 2-armed. In this
context the 4-armed pattern suggests a significant brightness of
the descending segments. The Perseus and Crux arms can be
partially identified with the descending segments of the ring
$R_2$.  Interestingly, the giant complex 475 ($l=352.8^\circ$,
$b=1.3^\circ$) \citep{russeil2007}, which is the brightest  in
the Crux arm and practically determines its position,  falls
exactly on the descending segment of the ring $R_2$ (see  its
location in Fig.~\ref{fit_sketch_AB}a).

We also studied the position of the outer rings with respect to
the tangential directions. It turned out that model of the
two-component outer ring can   also explain the appearance of
some of them: the line of sight in the direction of $l=284^\circ$
is almost tangential to the outer ring $R_2$, and the rays in the
directions of $l=310^\circ$ and of $51^\circ$ are tangents to the
ring  $R_1$ (Fig.~\ref{compare-I}). In addition, the lines of
sight in the range of $l=25\textrm{--}31^\circ$ are pointed  to
the end of the bar closest to the Sun. However, the directions of
$l=327^\circ$ and of $339^\circ$ cannot be identified with any
tangents to the rings or to the bar.

Fig.~\ref{compare-I} also exhibits  the gas-density distribution
in N-body model (Paper II). As was expected, the line of sight in
the directions of $l=284^\circ$ (Carina arm) and of $51^\circ$
(Sagittarius arm) cross a huge gas column at their way through
the combined $R_1R_2$ outer ring. The rays in the direction of
$l=25\textrm{--}31^\circ$ intersect a region of high gas content
located near the end of the bar. In distinction from models with
analytical bars, N-body model retains a lot of gas near the bar
ends.

\subsection{Ring $R_1R_2'$ and the diagrams ($l$, $V_{LSR}$)}

We assume  that the variations in the $^{12}$CO antenna
temperature  are caused by variations in the number of small
unresolved molecular cloudlets falling within the field of the
telescope \citep{mihalas1981}. If we associate these small clouds
with gas particles in our models then the most bright regions in
the observational maps must correspond to the regions of high
column density in the model diagrams.

\begin{figure*}
\centering \resizebox{14 cm}{!}{\includegraphics{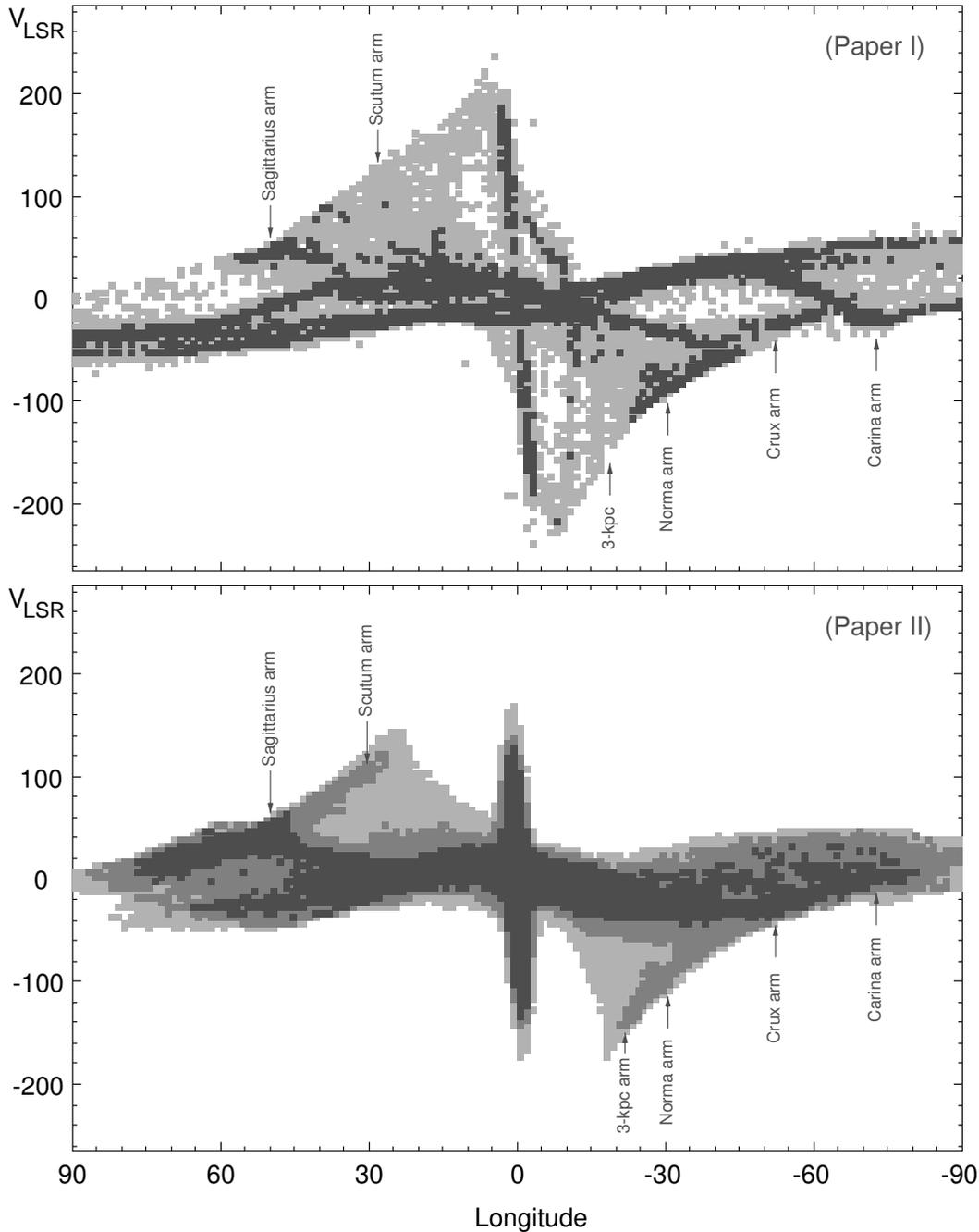}}
\caption{Diagrams (l, $V_{LSR}$) built for model with analytical
bar (Paper I) and for N-body simulation (Paper II). The darker
colors correspond to the cells with increasing number of
particles $n$. In the upper panel two gray tones represent cells
with $n$ below and above the average value $\overline{n}$
($\overline{n}=11$). In the lower panel three gray tones show the
cells with $10<n<\overline{n}$, $\overline{n}<n<2\overline{n}$,
and $n>2\overline{n}$ ($\overline{n}=427$). It also indicates the
maxima connected with the directions tangential to the spiral
arms.} \label{l-v-diagram}\end{figure*}

Fig.~\ref{l-v-diagram} shows the distribution of gas particles in
the plane (l, $V_{LSR}$) built for model with analytical bar
(Paper I) and for N-body simulation (Paper II). It also indicates
the positions of the observational maxima near the terminal
velocity curves which are supposed to correspond to the
directions tangential to the spiral arms. It is seen that the
model diagrams reproduce the intensity maxima in the direction of
the Carina, Crux, Norma, and Sagittarius arms. Moreover, N-body
model also creates the maxima in the directions of the Scutum and
3-kpc arms. Our models also produce the velocity peak of more
than $|V_{LSR}|>150$ km s$^{-1}$ in the central region,
$-5<l<5^\circ$.

\begin{figure*}
\resizebox{\hsize}{!}{\includegraphics{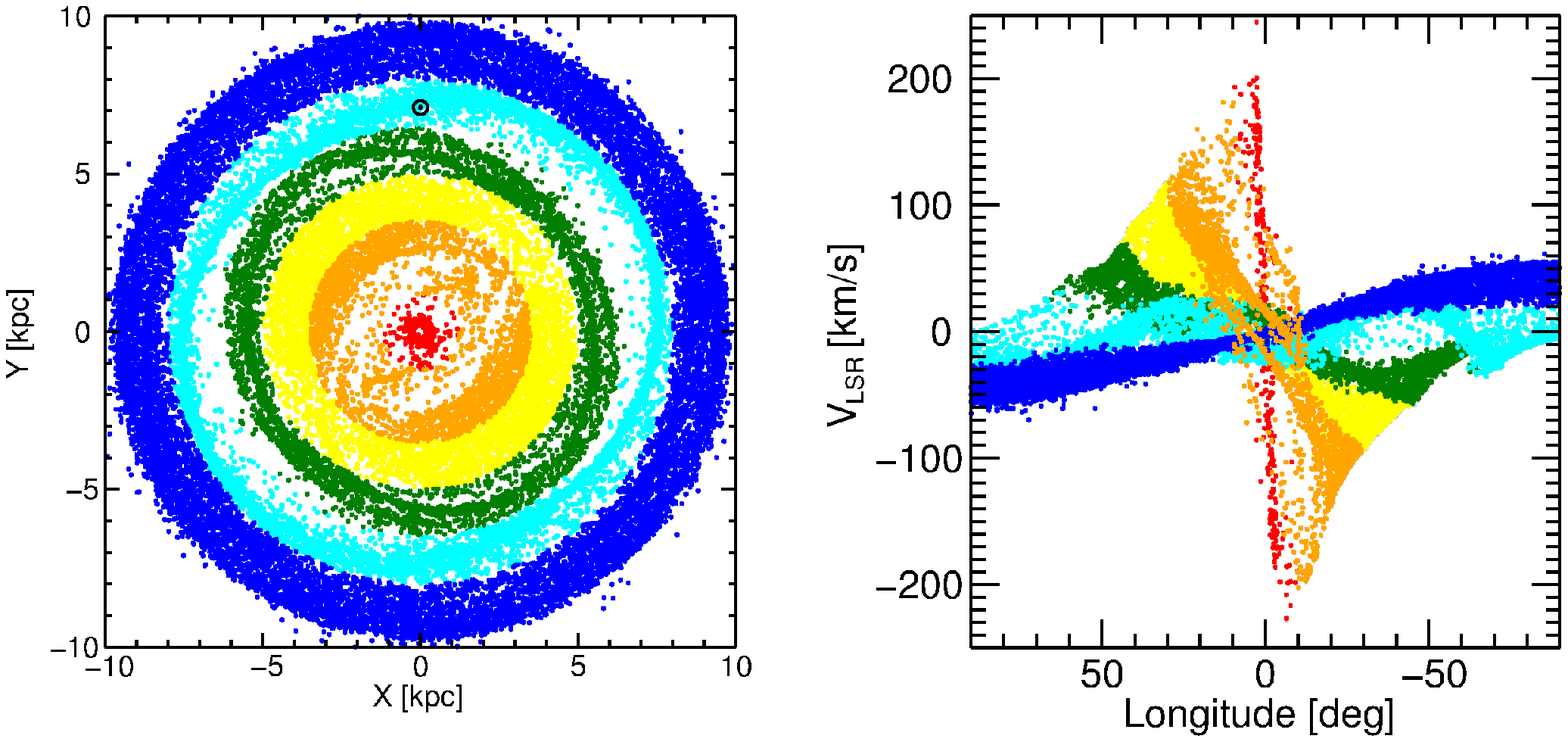}} \caption{ Left
panel: the distribution of gas particles in the Galactic plane in
model with analytical bar (Paper I). The size of the frame is 20
kpc. Particles located in different annuli are shown in different
gray tones. Right panel: the position of particles selected in
the diagram (l, $V_{LSR}$). The diagram has fan-shaped structure:
objects of different annuli are located in strip-like zones
turned at different angles to the vertical axis. The larger the
annulus, the greater the angle between the corresponding strip
and the vertical axis.} \label{fan}\end{figure*}

Fig.~\ref{fan} demonstrates the distribution of model particles
in the Galactic plane (Paper I) and their positions in the
diagram (l, $V_{LSR}$). The Galactic plane is divided into
annuli. The fan-shaped structure of the diagram is obvious:
particles located at different annuli occupy different strip-like
zones in the diagram.  The larger the radius of the annulus the
larger the angle between the corresponding strip and the vertical
axis. Interestingly,  the central peak is forming not only by
objects of the nuclear ring but also by particles  of the inner
ring.

Note that our model diagrams (l, $V_{LSR}$) don't reproduce the
so-called ``Molecular Ring''  on the observed CO-survey
(Fig.~\ref{dame}) -- the ridge of enhanced emission that extends
from the Scutum tangential point to the Norma one
\citep{dame2001}. This observational feature is sometimes
interpreted  as a molecular ring \citep{binney1991} or as spiral
arms  emanating from the bar \citep{fux1999}. In any case our
models need some modification to keep more gas near the bar ends.


\subsection{Ring model of the Galaxy}

\begin{figure}
\centering \resizebox{5 cm}{!}{\includegraphics{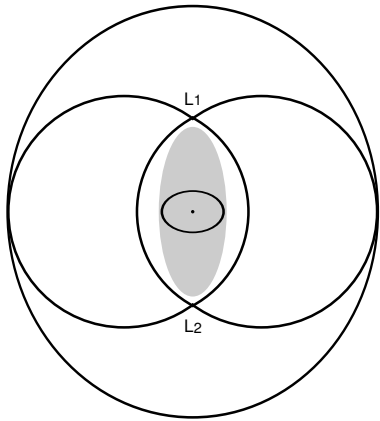}}
\caption{Basic model of the galactic ring structure. It includes
the bar (gray ellipse), the nuclear ring which is represented by
an ellipse aligned perpendicular to the bar, the inner ring
elongated along the bar, the ``8''-shaped outer ring $R_1$
stretched perpendicular to the bar, and the outer ring $R_2$
aligned with the bar.} \label{basic_sketch}\end{figure}

Let us consider   a new model of the Galaxy that includes two
outer rings, the inner ring, and the nuclear ring.
Fig.~\ref{basic_sketch} represents the basic sketch of the
galactic ring structure  composed on the base of sketches
designed by \citet{buta1986}.  The resonance rings are supported
by the periodic orbits which are elongated parallel or
perpendicular to the bar and change their orientation near the
OLR, CR, and ILR(s),  some chaotic orbits also share similar
characteristics \citep{contopoulos1980,contopoulos1989}. But not
only resonant processes determine the formation of  ring-like
structures in the galactic disks. They are also affected by the
gas flow outwards or inwards due to torque from the bar.  The
central region often includes the inner spiral arms that connect
Lagrangian points $L_1$ and $L_2$ with the nuclear ring or with
the galactic center, which are shocks caused by the bar
\citep{athanassoula1992}. Note also that  outer rings $R_1$ and
$R_2$ in the basic sketch connect with each other, but such
connection is sometimes absent in  numerical simulations and in
images of real galaxies (for example, NGC 1211). The connection
between the inner ring and the outer ring $R_1$ can be missing as
well \citep[for example, NGC 3081,][]{buta2007}

\begin{figure*}
\resizebox{\hsize}{!}{\includegraphics{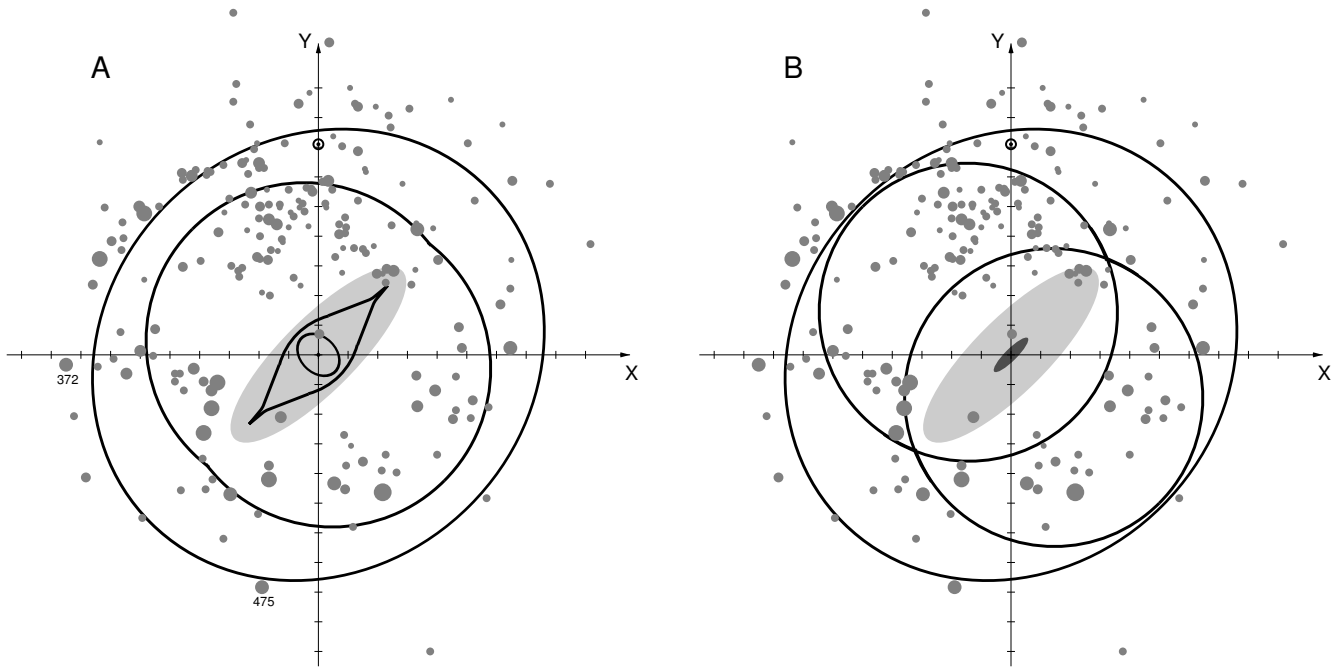}} \caption{Ring
structure applied  to the Galaxy.  In both sketches the bar is
represented as a gray ellipse with the semi-axes $a=4.0$ and
$b=1.2$ kpc. The position angle of the Sun with respect to the
bar is $\theta_b=45^\circ$. The outer ring $R_2$ is shown by  an
ellipse elongated along the bar with the semi-axes $a_2=8.0$ and
$b_2=7.2$ kpc.   Sketch A is determined by the distribution of
particles in model with analytical bar (Paper I) while sketch B
is based on the distribution of  particles in N-body simulations
(Paper II).  We can see a gap between two outer rings $R_1$ and
$R_2$ in sketch A but it is absent in sketch B. Sketch A has the
more elongated and smaller inner ring in comparison with that in
sketch B. There are also some differences in the shape and
orientation of the nuclear gas condensation  in sketches A and B.
Also shown the distribution of giant star-forming complexes
($U>60$ pc cm$^{-2}$) \citep{russeil2007}. The value of $R_0$ is
adopted to be $R_0=7.1$ kpc} \label{fit_sketch_AB}\end{figure*}

The application of  the basic ring structure to the Galaxy
doesn't give an unambiguous picture.  On the basis of numerical
simulations we designed two sketches of the Galactic spiral
structure (Fig.~\ref{fit_sketch_AB}). In sketch A we try to
reproduce the distribution of gas particles in model with
analytical bar (Paper I) while in sketch B -- the distribution of
gas and star particles  in N-body simulation (Paper II). Both
sketches have many similar features: the bar is represented as an
gray ellipse with the semi-axes $a=4.0$ and $b=1.2$ kpc, the
position angle of the Sun with respect to the bar equals
$\theta_b=45^\circ$, the outer ring $R_2$ is approximated by an
ellipse elongated along the bar with the semi-axes $a_2=8.0$ and
$b_2=7.2$ kpc, which is in good agreement with the distribution
of OB-particles in models with analytical bars (Paper I). The
main differences of the sketches A and B lie in the size of the
ring $R_1$, in the shape of the inner ring, and in the
orientation of the central gas condensation.

In sketch A  the CR of the bar lies   at $R=4.0$ kpc -- just at
the bar ends. The $R_1$ ring reaches only the radius of $R=6.0$
kpc thereby forming a gap  between  the two outer rings. The
inner structure is represented by the pointed inner ring
connecting the bar ends with the nuclear ring. The connection
between the inner  ring  and the outer ring $R_1$ is also absent.
The nuclear ring in represented by an ellipse elongated
perpendicular to the bar with the semi-axes $a_n=0.8$ and
$b_n=0.6$ kpc.

In sketch B the CR of the bar is located at $R=4.6$ kpc. The
$R_1$ ring begins near the CR and reaches for the OLR of the bar
so that there is no gap between the rings $R_1$ and $R_2$. In
N-body simulation the ring $R_1$ is forming mainly in stellar
population. The gaseous inner ring has more round shape here
compared with sketch A. The central gas condensation is
represented by an ellipse elongated along the bar with semi-axes
$a_n=0.8$ and $b_n=0.2$ kpc.

Both sketches can easily explain the location of the
Sagittarius-Carina arm with respect to the Sun. This arm can
consist of two ascending segments of the outer rings $R_1$ and
$R_2$. At the first glance  objects of the Carina arm are forming
more open structure, but this impression is mainly based on the
position of the  complex 372 ($l=311.2^\circ$, $b=-0.4^\circ$)
\citep{russeil2007}(Fig.~\ref{fit_sketch_AB}a). However, we can
move it along the line of sight  so that it falls exactly on the
ring $R_2$  (for more details see section 5.4).

The mid-infrared observations show an excess of old stars  in the
direction of the Centaurus ($l\approx -50^\circ$) and Scutum
($l\approx 25\textrm{--}31^\circ$) arms but not  in the direction
of the Sagittarius arm ($l\approx +50^\circ$)
\citep{drimmel2000,churchwell2009}. Sketch A can easily explain
an increase of the density of old stars in the direction of the
Centaurus arm (another name of the Crux arm). The line of sight
in the direction of $l\approx -50$ is nearly tangent to the outer
ring $R_1$ (Fig.~\ref{compare-I}). Observations and modelling
show that  the $R_1$ rings can be forming in the stellar
subsystem, but the $R_2$ rings usually appear only in gas
component \citep{byrd1994,rautiainen2000}. However, we cannot
explain the absence of an excess of old stars in the direction of
the Sagittarius arm   -- in our model the Centaurus ($l\approx
-50$) and Sagittarius ($l\approx +50$) arms are the segments of
the same ring $R_1$ and, consequently,  must have the same
nature.

Distribution of optical objects  in the Galactic plane also gives
some evidences of the existence of a gap between the Sagittarius
and Carina regions. \citet{humphreys1979}  shows that
OB-associations and young open clusters concentrate either in the
Sagittarius region or in the Carina one but not in between.
Recent studies based on the analysis of the distribution of young
open clusters \citep{dias2002,mermilliod2003} and classical
long-period Cepheids \citep{berdnikov2000} confirm the presence
of the gap in the distribution of young objects along the
Sagittarius-Carina arm \citep{majaess2009}. This fact needs very
accurate interpretation because  spiral arms can have patchy
structure. On the other hand, different kinematics of these
regions suggests that they  can belong to the different outer
rings (Paper I).

Note that the  inner ring in sketch B is larger and less
elongated   than that in sketch A (Fig.~\ref{fit_sketch_AB}).
Probably, this larger ring can correspond to the case where the
inner rings are forming farther away from the bar
\citep{grouchy2010}. Nevertheless, both types of the inner rings
can be associated with the 3-kpc arm and its counterpart
\citep{fux1999,dame2008,rodriguez-fernandez2008}.

At the moment we cannot say which conditions determine the exact
place and shape of the inner ring-like structures and those of
the outer rings $R_1$ in sketches A and B. In principle, the
difference between them   may  be related to different kinds of
orbits creating  them: the very pointed inner ring in sketch A
could be formed by the ``classical'' orbits that are found in
barred potentials \citep{contopoulos1989} whereas the
rings/pseudorings in sketch B could be formed by manifold orbits
\citep{athanassoula2010}.

Our sketches  also exhibit conspicuous differences in the  shape
and orientation of the nuclear gas condensation: in sketch A it
is more round and elongated perpendicular to the bar while in
sketch B it is oriented along the bar and looks   like a
secondary bar \citep[][and references therein]{erwin2011}. All
our models have two ILR resonances located at distances
$R_{ILR}=0.2$ and $1.5$ kpc (Papers I and II), so the difference
between them cannot be caused by their position. Probably, it
appears due to some features of the gas inflow. Special numerical
simulations of the gas flow in the central region of the Galaxy
show that 1-kpc nuclear ring can be holed and contain additional
elliptical gas condensation with semi-axis of $a\approx200$ and
$b\approx100$ pc which is associated with the Central Molecular
Zone (CMZ) \citep{ferriere2008,rodriguez-fernandez2008}, but our
models have not enough resolution to reproduce this detail.


\subsection{Kinematical distances}

Most giant star-forming complexes have only kinematical distances
which were calculated from  the kinematical models with purely
circular rotation law. \citet{russeil2003} reckons that
photometrical distances for stars exciting HII regions are
determined with the errors $20\textrm{--}30$\%. The errors in
kinematical distances depend on the direction, but, on average,
the deviations of the  velocity $V_{LSR}$ from the rotation curve
of 15 km s$^{-1}$ correspond to the error of $\sim20$\% in
kinematical distances.  However, this estimation was derived
under the assumption that we always made a correct choice between
the ``far'' and ``near'' distances on the same line of sight, but
the non-circular gas motions significantly complicate this
choice. In the case of wrong choice the distance error can exceed
100\%.

We compared the observed $V_{LSR}$ velocities of giant star
forming complexes from the catalog by \citet{russeil2007} with
the model velocities of gas particles in model  with analytical
bar (Paper I). For each complex we selected model particles
located within 200 pc from the observed position of a complex
(l,~r) and calculated their mean velocity along the line of
sight. The mean difference between the model and observed
$V_{LSR}$ velocities is found to be $\Delta V=16$ km s$^{-1}$
which doesn't exceed significantly the mean difference between
the observed $V_{LSR}$ velocity and velocity calculated from the
model rotation curve $\Delta V=11$ km s$^{-1}$. Formally, the
kinematical distances by \citet{russeil2007} are quite
reasonable.

The  scale of kinematical distances is determined by the distance
scale of objects used for calculation of rotation curve. If
distances for objects studied and rotation curve are
self-consistent then the velocity deviations from the rotation
curve are always minimal and   practically independent of the
distance scale chosen.

\begin{figure}
\resizebox{\hsize}{!}{\includegraphics{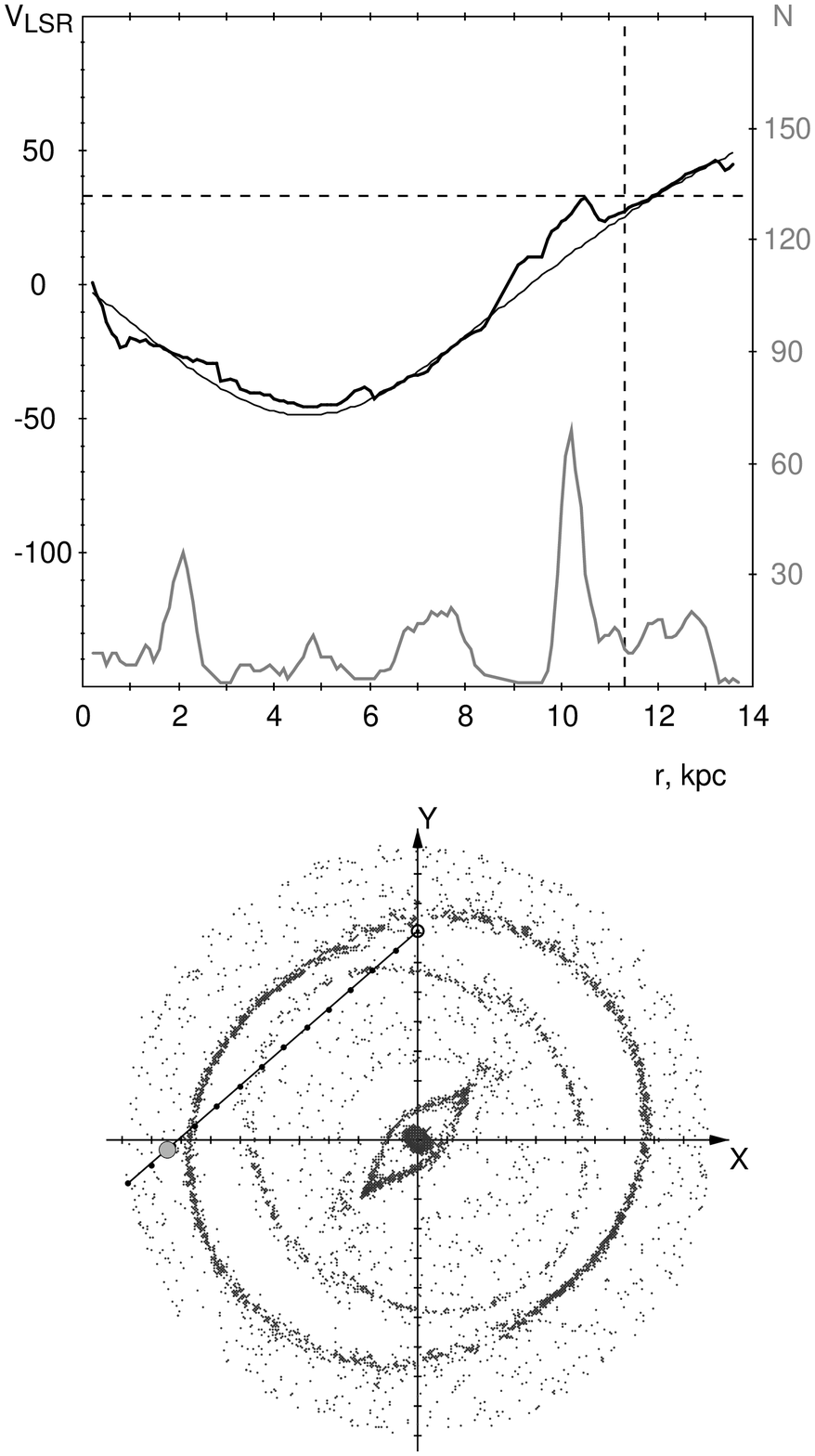}} \caption{{\it
Upper plot}: Model $V_{LSR}$ velocities (black thick line)
calculated for different heliocentric distances $r$ of the
star-forming complex 372  ($l=311.2^\circ$, $b=-0.4^\circ$)
\citep{russeil2007}. Dashed lines indicate the observed $V_{LSR}$
velocity ($V_{LSR}=33.0$ km s$^{-1}$) and the catalog value of
the kinematical distance reduced to  the short distance scale
($r=11.3$ kpc). Gray line indicates the number of particles $N$
within the 200-pc circle at the chosen $r$. Left vertical axis
(black one) shows the scale for $V_{LSR}$ velocities while the
right axis (gray one) exhibits the scale for $N$. Thin black line
indicates $V_{LSR}$ calculated through the model rotation curve.
{\it Lower plot}: the position of the complex 372 in the galactic
plane with respect to the distribution of model particles (Paper
I).  The line of sight with 1-kpc divisions is also shown.}
\label{VLSR}\end{figure}

Fig.~\ref{VLSR} shows model $V_{LSR}$ velocities  calculated for
different heliocentric distances $r$ of the  star-forming complex
372 ($l=311.2^\circ$, $b=-0.4^\circ$) in the catalog by
\citet{russeil2007}. We selected model particles (gas and OB)
located within 200 pc from the chosen position of the complex and
calculated their model $V_{LSR}$ velocity. The number of model
particles $N$ within the 200-pc circle is also shown. The
positions of the rings correspond to the maxima on curve $N(r)$.
For each $r$ we also determine the $V_{LSR}$ velocity through the
model rotation curve. We can see that complex 372 can be moved
from the distance $r=11.3$ to $10.2$ kpc to fall exactly on the
ring $R_2$, and its new position is in a good agreement  with the
observed $V_{LSR}$.

\section{Conclusions}

Model of the Galaxy with the   outer ring $R_1R_2'$ can explain
some large-scale morphological features of the Galactic spiral
structure. Ascending segments of the rings can be regarded as
fragments of the trailing spiral arms while descending ones -- as
fragments of the leading arms. We found that the Carina arm falls
well on the ascending segment of the  ring $R_2$. Note also that
the objects of the Sagittarius arm are located near the ascending
segment of the  ring $R_1$. The Carina-Sagittarius arm can
consist of two ascending segments of the   outer rings $R_1$ and
$R_2$, which almost touch each other near the Carina region. It
is possible that another pair of ascending segments of the outer
rings can be identified with the Norma-Cygnus arm symmetrical to
the Carina-Sagittarius one. The Perseus and Crux arms can be
partially identified with the descending segments of the ring
$R_2$. Thus, the two-component outer ring $R_1R_2'$ can be
mistakenly interpreted as the 4-armed spiral pattern.

Fourier analysis of the distribution of OB-associations with the
same kinematic characteristics over spiral harmonics shows the
presence of a leading component in the spiral structure of the
Galaxy \citep{melnik2005}.  The sample includes OB-associations
whose radial component $V_R$ of  velocity  is directed toward the
Galactic center. The appearance of the leading spiral agrees with
the position of the Sun near the descending segment of the ring
$R_2$, which can be thought as a fragment of the leading spiral
arm.

Model of the two-component  outer ring could also explain the
existence of some tangential directions corresponding to the
emission maxima near the terminal velocity curves. Model diagrams
(l, $V_{LSR}$) reproduce the maxima in  the direction of the
Carina, Crux, Norma, and Sagittarius arms. Additionally, N-body
model yields the maxima in the directions of the Scutum and 3-kpc
arms.

 On the basis of numerical simulations we propose two sketches
of the  ring structure of the Galaxy which includes  the bar, two
outer rings, the inner ring, and the nuclear gas condensation
forming the nuclear ring and/or the secondary bar
(Fig.~\ref{fit_sketch_AB}ab). Both sketches can explain the
position  of the Carina-Sagittarius arm with respect to the Sun.
Sketch A can also explain the existence of an excess of old stars
in the direction of the Centaurus arm $l\approx -50$.

\section*{\rm \large ACKNOWLEDGMENTS}

We wish to thank Heikki Salo for using his simulation code.  We
are grateful to N.Ya. Sotnikova and P. Grosbol for useful remarks
and interesting discussion. This work was partly supported by the
Russian Foundation for Basic Research (project
nos.~10\mbox{-}02\mbox{-}00489).

\end{document}